\documentclass[12pt,a4paper,oneside]{article}
\usepackage{amsmath}

\begin{document}
\begin{center}

           CASIMIR SURFACE FORCE ON A DILUTE DIELECTRIC BALL\\[0.5cm]

                  Iver Brevik\footnote{Corresponding author.
E-mail address: iver.h.brevik@maskin.ntnu.no}\\
                    Division of Applied Mechanics\\
            Norwegian University of Science and Technology\\
                       N- 7034 Trondheim, Norway\\
\bigskip
and\\
\bigskip
                     Valery Marachevsky\\
                   Department of Theoretical Physics\\
                       St. Petersburg University\\
                    198 904 St. Petersburg, Russia\\[0.5cm]
\end{center}
\bigskip
\begin{center}

\bigskip
January 1999\\
\bigskip

\end{center}

\begin{abstract}
    The Casimir surface force density $F$ on a dielectric dilute spherical ball
of
 radius $a$, surrounded by a vacuum, is calculated at zero temperature. We treat
 $(n-1)$ ($n$ being the refractive index) as a small
parameter. The dispersive properties of the material are taken into account by
adopting a simple
dispersion relation, involving a sharp high frequency cutoff at
$\omega=\omega_0$. For a {\it nondispersive}
medium there appears (after regularization) a finite, physical, force
$F^{nondisp}$ which is repulsive.
By means of a uniform asymptotic expansion of the Riccati-Bessel functions we
calculate $F^{nondisp}$ up
to the fourth order in $1/\nu $. For a {\it dispersive} medium the main part of
the force $F^{disp}$ is
also repulsive.   The dominant term in $F^{disp}$ is proportional to
$(n-1)^2 \omega_0^3/a $ , and will under usual physical
conditions outweigh $F^{nondisp}$ by several orders of magnitude.
\end{abstract}

\newpage

\section{Introduction}
    Consider a compact dielectric nonmagnetic ball of radius $a$, surrounded by
a vacuum.
The permittivity of the material is  $ \varepsilon=n^2$ , $n$ being the
refractive index.
 The purpose of the present work is
to calculate the "ordinary" Casimir surface density $F$ on the sphere (i.e. with
the exception of the
electrostrictive force), at zero temperature, making use of the same
calculational technique as in an
earlier  work \cite{brevik94}.. For mathematically simplifying reasons we will
be concerned only with the case of
dilute media, meaning that $n - 1 \ll 1$.  As in \cite{brevik94}, we will make
use of the uniform asymptotic
expansion (also called the Debye expansion) for the
 modified  Bessel functions \cite{abramowitz64}. Here $1/\nu$  is the expansion
parameter, with
$\nu  =  l + 1/2 $ , $ l  = 1, 2, 3,.....$.
In the main body of this paper we will expand the formalism to second order in
$(n - 1)$, and to fourth
order in $1/{\nu} $ .  In  Appendix A  we carry out the expansion in $(n - 1)$
three orders further, to the fifth
order, at the price, however, of retaining the Debye expansion to second order
in $1/\nu$ . This treatment
improves considerably on the theory given in \cite{brevik94}. We use the
mathematical package Maple V  Release
4 Power Edition. It turns out to be necessary to keep relatively high accuracy
in this problem, since
subsequent terms in the uniform asymptotic expansion alternate in sign. Thus, if
one simply truncates the series
after a few terms, without observing this alternating effect, one risks getting
even the sign in the final
force wrong.

    The ultimate goal in a calculation of this type is to get a Casimir force
expression that is directly
comparable to experiment; if  not a real experiment, at least a Gedanken
experiment. To achieve this
goal, one must observe that there are several contributions to the surface
force. Let us try to put the
things into perspective, by writing down the general expression for the
electromagnetic volume force
density $ {\bf f} $ in a charge - free nonmagnetic medium (cf., for instance,
the book \cite{landau84}
 or the review \cite{brevik79})
\begin{equation}
{\bf f}= -\frac{1}{2}E^2\nabla \varepsilon + \frac{1}{2}\nabla\left[E^2\rho
\left(\frac{\partial \varepsilon}{\partial \rho}\right)_T\right] +
\frac{\varepsilon -1}{c}\frac{\partial}{\partial t}({\bf E}\times {\bf H}).
\end{equation}

Here $\rho$    is the mass density of the fluid.

    Consider the first term in (1). We may call it the Abraham - Minkowski
force. It is equal to zero
in the interior homogeneous region of the medium and acts in the boundary region
around $r = a$ only.
>From thermodynamic perturbation theory we know in general that  $\varepsilon  >
1$ for a dielectric at zero
frequency \cite{landau84}, Sect.14, and we expect the same to be true for
moderate and high frequencies also, except
possibly from special frequency bands. That is, since the sphere is surrounded
by a vacuum, we would
expect the first term in (1) to give a repulsive surface force. This turns out
to be true also, in classical
theory. In quantum theory, however, we cannot be sure in advance about the
direction of the force; the
result depends on the magnitude of the contact term being subtracted off in the
regularization procedure. One
has to carry out the calculation in detail in order to determine even the sign.

    The Abraham - Minkowski force can in turn be divided into two parts:

(1) A nondispersive part. It corresponds to letting the ultraviolet
(nondimensional) cutoff frequency
   $x_0$ go to  infinity. The early works on the Casimir effect took into
account this part of the force
 only. The nondispersive force is formally divergent, but can usually be easily
regularized. The  recipe
 is to (i) put $x_0 =\infty$  ; (ii) evaluate the remaining divergent sum over
$l$  , from $l  = 1$
 to $l  = \infty$, by the Riemann zeta - function method. We stressed the
efficiency of this recipe in \cite{brevik94};
 cf. also some other related prior papers \cite{milton83} - \cite{brevik90}. The
method gives, in a simple way,
 results that are in
 agreement with what can be obtained in other ways. The quantum nondispersive
force turns out in our
calculation to be repulsive, thus qualitatively in agreement with the classical
result noted above.

(2) There is a dispersive part of the Abraham - Minkowski force, caused by the
cutoff $x_0$ . In an earlier
work on a spherical shell \cite{brevik90} we actually found the dispersive force
to be attractive. The   dispersive
 force is under usual physical conditions {\it stronger} than the nondispersive
force. Candelas \cite{candelas82}
seems to have been the first to emphasize the importance of the dispersive
force. His general ideas were
later on essentially supported by explicit model calculations \cite{brevik88},
\cite{brevik89}.

    Consider next the second term in (1). This is the {\it electrostriction}
force. It generally acts inwards.
Whether this force is detectable in a real experiment or not, depends on the
detailed circumstances.
Usually, there is established a compensating force from the elastic pressure in
the interior of the fluid,
so that the effect from electrostriction does not show up explicitly in the
measurements. Only under
special conditions, such as in the Goetz - Zahn non-equilibrium experiment
testing the attractive force
between two condenser plates immersed in a dielectric liquid when there is a
high-frequency electric
field between the plates \cite{goetz58}, \cite{zahn62}, does one have the
possibility to measure the electrostriction force. The
point is that the electromagnetic force will then vary so quickly that the
elastic pressure does not have
sufficient time to build itself up. What determines the time scale here, is
obviously the transit time for
sound waves in the body. (The Goetz - Zahn experiment, as well as some other
related experiments, are
discussed in detail in \cite{brevik79}.)
    In (1), we have deliberately written $(\partial\varepsilon/\partial \rho)_T
$ as an isothermal partial
 derivative. This is often
appropriate in practical cases, but is an assumption that has to be considered
with some care since if the
deformation of the body is occurring rapidly it will be more appropriate to take
the adiabatic derivative
$ (\partial \varepsilon/\partial \rho)_S $    instead. If the fluid is non-
polar, as one can usually assume
 at high frequencies, then one need
not distinguish between the isothermal and adiabatic cases since the
permittivity depends on density
only, though the Clausius - Mossotti relation. The electrostrictive contribution
to the Casimir effect has
not been paid much attention to in the past. There is a paper by one of us some
years ago \cite{brevik82}, and
recently Milton and Ng have returned to the topic \cite{milton97}. In the
present case, where we are considering
the Casimir force on a static spherical boundary, it is clear that a
compensating elastic pressure will be
built up on the inside. Because of this, most possibilities for measuring the
surface force that one may
conceive of, would not be able to detect the electrostrictive contribution at
all. A measurement of
electrostriction would require a measurement of the local pressure in the
interior region.

    Consider finally the third term in (1). This is the Abraham term, detectable
under special
circumstances at low frequencies \cite{brevik79}, but fluctuating out at higher
frequencies, especially at optical
frequencies. In our case, the Abraham term does not contribute. It fluctuates
out.

    Summing up so far, we see that it is the Abraham - Minkowski term in (1)
that gives the most
important contribution to the surface force. We will be concerned with this term
in the following, and
will consider both subclasses (1) and (2) listed above. In the next section we
highlight the general
Green-function formalism, and explain the regularization method that we use.
Section 3 is devoted to
a study of the nondispersive force. By using the expansion technique mentioned
above we derive the
expression for $F^{nondisp}$, to order $1/\nu^4$, in (27). This force is
repulsive.
Some earlier results,
derived by purely analytical means, are corrected. As regards the dispersive
force, considered in Section
4, we restrict ourselves to giving essentially estimates. For high ultraviolet
nondimensional cutoff  $x_0$
the dispersive force is given as the sum of expressions (39) and (46). The
dispersive force for a
compact ball is strong, and it is repulsive. It may be as large as about $10^6 -
10^9$ times the nondispersive
force. This fact does not seem to have been universally recognized before.

    We put henceforth $\hbar  = c = 1$. Electromagnetic Heaviside - Lorentz
units are employed.

\section{Basic formalism. Regularization}

    The general formalism was developed in our earlier paper [1], but for the
benefit of the  reader
we will recapitulate some of the essential points.

    The dispersion of the material is accounted for in a simple way, by taking
the
permittivity $\varepsilon(i \hat\omega)$ as a function of imaginary frequencies
$\hat\omega$
to be a step function:
  $\varepsilon(i \hat\omega)= n^2=constant >1$ for $\hat\omega < \omega_0$,
$\varepsilon(i \hat\omega)=1$ for $\hat\omega > \omega_0$.  Thus there is one
single "absorption" frequency
$\omega_0$,  serving as an
ultraviolet cutoff.

    When considering the two-point functions, we let the two spatial points $r$
and $r'$  be separated
in the radial direction. There are two scalar Green functions in the problem,
$F_l(r,r')$ and
$G_l(r,r')$, defined by
\begin{equation}
r, r'< a: ~~~F_l,G_l = inkj_l(nkr_<)[h_l^{(1)}(nkr_>)-\tilde
A_{F,G}(ka)j_l(nkr_>)],
\end{equation}
\begin{equation}
r,r'>a:~~~F_l,G_l = ik\left[j_l(kr_<)-\tilde B_{F,G}(ka)
h_l^{(1)}(kr_<)\right]h_l^{(1)}(kr_>).
\end{equation}
Here $k = |\omega |$,  $j_l$   is the spherical Bessel function, and $h_l^{(1)}
$  the spherical Hankel function of
 the first kind. The coefficients $ \tilde A_{F,G} $   and $\tilde B_{F,G} $
are related to the boundary.

    We shall work with the two-point functions that contain the surface  -
induced contributions
only, and thus have to subtract off the contact terms corresponding to a uniform
medium. It means that
the two-point functions refer to "disturbed" quantities caused by the
boundaries. Specifically, we let
 (i) the contact force on the inner side $  r = a- $   be calculated as if the
inner medium be filling all space;
(ii) the contact force on the outer side  $ r = a+ $  be calculated as if the
outer medium be filling all space.
This means that the first term in (2) has to be subtracted off as a contact
term. Similarly the first term
in (3) has to be subtracted off.

We stress that this method of regularization is not an arbitrary choice. It is
the only kind of
regularization that separates off the volume terms, and it is the only method
that permits us to write the relation between the surface force $F$ and the
Casimir energy $E$ as $F=-(1/4 \pi a^2)\partial E/\partial a $. It implies, in
particular, that both the surface force and the Casimir energy goes to zero when
the radius goes to infinity, what
 seems to us a most natural result. Our method of regularization is the same as
used by Milton \cite{milton80},
\cite{milton96}, \cite{milton99}; cf. also \cite{milton97}, \cite{brevik98}. It
should be mentioned, however,
that this method is different from the one advocated recently by Carlson {\it et
al.} \cite{carlson97a},
\cite{carlson97b}.  We shall call the modified two-point functions,
constructed by the boundary-related parts of the Green functions only, the
"effective"two-point
functions, and write them as $ <  >_ {eff} $. We introduce the nondimensional
cutoff frequency
\begin{equation}
x_0= \omega_0 a ,
\end{equation}
and perform a complex frequency rotation
\begin{equation}
k\rightarrow i \hat k = i \hat\omega,~~~ka \rightarrow i \hat\omega a \equiv i
x,
\end{equation}
whereby the Riccati - Bessel functions $ s_l $   and  $ e_l $  defined by
\begin{equation}
s_l(x)= \sqrt{\frac{\pi x}{2}}I_{\nu}(x),~~~~e_l(x)=
\sqrt{\frac{2x}{\pi}}K_{\nu}(x)
\end{equation}
correspond to the Wronskian  $ W\{s_l,e_l\}=-1 $.  Here $\nu=l+1/2$,   $I_{\nu}$
and
$K_{\nu}$    denoting modified
Bessel functions.

The effective electric field products in the limit $r' \rightarrow r$  in the
interior region $ r<a$   are
\begin{equation}
\langle E_r(r) E_r(r')\rangle_{eff}=-\frac{1}{\pi r^4}\int_0^{x_0}
\frac{dx}{n^3 x}\sum_{l=1}^{\infty}\frac{2l+1}{4 \pi}l(l+1)A_G(x)s_l^2(n x
\tilde r),
\end{equation}
\begin{equation}
\langle E_\perp(r)E_\perp(r')\rangle_{eff} = \frac{1}{\pi r^2
a^2}\int_0^{x_0}\frac{xdx}{n}
\sum_{l=1}^{\infty}\frac{2l+1}{4 \pi}\{A_F(x)s_l^2(nx\tilde r)
-A_G(x) [s_l(nx\tilde r) ]^2 \} ,
\end{equation}
with $\tilde r=r/a $,  $ n= n(ix)$, prime meaning differentation with respect to
the whole argument. We  need
the expressions for the frequency-rotated coefficients:
\begin{equation}
A_F(x)=\frac{e_l(nx)e_l\prime (x)-ne_l(x)e_l\prime (nx)}{s_l(nx)e_l\prime (x)-
ne_l(x)s_l\prime(nx)},
\end{equation}
\begin{equation}
A_G(x)=\frac{ne_l(nx)e_l\prime(x)-e_l(x)e_l\prime(nx)}{ns_l(nx)e_l\prime(x)-
e_l(x)s_l\prime(nx)}
\end{equation}
(it may be noted that $ \tilde A_{F,G}(ix)=(-1)^{l+1}A_{F,G}(x)) $.  Similarly
in the exterior region
$ r>a$ we have
\begin{equation}
\langle E_r(r) E_r(r')\rangle_{eff}= -\frac{1}{\pi
r^4}\int_0^{x_0}\frac{dx}{x}\sum_{l=1}^{\infty}
\frac{2l+1}{4\pi}l(l+1)B_G(x)e_l^2(x \tilde r),
\end{equation}
\begin{equation}
\langle E_{\perp}(r)E_{\perp}(r')\rangle_{eff} = \frac{1}{\pi r^2 a^2}
\int_{0}^{x_0}  xdx
\sum_{l=1}^{\infty}
\frac{2l+1}{4 \pi}\{B_F(x)e_l^2(x \tilde r)
- B_G(x)[e_l\prime(x \tilde r)]^2 \},
\end{equation}
with
\begin{equation}
B_F(x)=\frac{s_l(nx)s_l\prime(x)-n s_l(x) s_l\prime(nx)}{s_l(nx)e_l\prime(x)-
ne_l(x)s_l\prime(nx)},
\end{equation}
\begin{equation}
B_G(x)=\frac{ns_l(nx)s_l \prime(x)-s_l(x)s_l \prime(nx)}{ns_l(nx)e_l \prime(x)-
e_l(x)s_l \prime(nx)}.
\end{equation}
The surface force density, as calculated by use of the Maxwell stress tensor, is
$ F = F_{int} + F_{ext} $ , where the
interior force is
\begin{equation}
F_{int}=\frac{1}{2 \pi a^4}\int_0^{x_0}nxdx\sum_{l=1}^{\infty}\frac{2l+1}{4 \pi}
\Lambda_l^{int}\left[\ln (\frac{s_l\prime(nx)}{s_l(nx)})\right]^{\prime},
\end{equation}
with
\begin{equation}
\Lambda_l^{int}=[A_F(x)+A_G(x)]s_l(nx)s_l\prime(nx).
\end{equation}
Similarly the exterior force is
\begin{equation}
F_{ext}=-\frac{1}{2 \pi a^4}\int_0^{x_0}xdx\sum_{l=1}^{\infty}\frac{2l+1}{4 \pi}
\Lambda_l^{ext}\left[ \ln ( \frac{-e_l\prime(x)}{e_l(x)}) \right]^{ \prime},
\end{equation}
with
\begin{equation}
\Lambda_l^{ext}=[B_F(x)+B_G(x)]e_l(x)e_l\prime(x).
\end{equation}
These expressions assume the step-function dispersion relation for $ n(ix)$, as
explained above. The value
of the constant $n$ for $ x < x_0 $ is however at this stage arbitrary; $(n-1)$
is not necessarily small.

The natural question is now: are the force expressions (15) and (17) finite? The
answer is no. This can be seen by means of the uniform asymptotic series to be
employed below, or by direct machine calculation. Thus regularization is called
for. The regularization procedure is quite straightforward, at least for the
case of dilute media, the case to which we now turn.

\section{Nondispersive force}
     We assume henceforth $(n-1)$ to be a small quantity, and expand the
formalism, by
means of Maple, up to order $(n-1)^2$. We make use of the uniform asymptotic
expansions for the Riccati-Bessel functions, as explained in \cite{brevik94}. As
mentioned above,
we insert these expansions up to order $1/\nu^4$  .

     As we mentioned in  Sect.1, the nondispersive case can be treated by
setting $x_0=\infty$.   We
introduce the symbols
\begin{equation}
z=x/\nu,~~~~t(z)=(1+z^2)^{-1/2},
\end{equation}
and calculate in the interior region the integrand of Eq.(15), using the Debye
expansion. Taking the
integral from $z=0$ to $z=\infty$ we obtain
\begin{eqnarray}
\int_0^{\infty}dz n\nu^3z \Lambda_l^{int}\left[\ln (\frac{s_l
\prime(nx)}{s_l(nx)})\right]
^{\prime} \nonumber \\
=-\frac{(n-1)\pi\nu^2}{8}\left[1-
\frac{41}{128}\frac{1}{\nu^2}+\frac{711}{16384}\frac{1}{\nu^4}\right] \nonumber
\\
+\frac{5(n-1)^2\pi\nu^2}{32}\left[1-\frac{592}{525}\frac{1}{\pi\nu}-
\frac{229}{640}\frac{1}{\nu^2}
+\frac{29504}{75075}\frac{1}{\pi\nu^3}
+\frac{3771}{81920}\frac{1}{\nu^4}\right].
\end{eqnarray}
This is to be inserted into the expression for the interior nondispersive
surface force density
 $F_{int}^{nondisp} $  which, according to (15), can be written as
\begin{equation}
F_{int}^{nondisp}=\frac{1}{4 \pi^2
a^4}\sum_{l=1}^{\infty}\int_0^{\infty}dzn\nu^3z\Lambda_l^{int}
\left[\ln(\frac{s_l\prime(nx)}{s_l(nx)})\right]^{\prime}.
\end{equation}
In the exterior region we calculate similarly
\begin{eqnarray}
-\int_0^{\infty}dz\nu^3
z\Lambda_l^{ext}\left[\ln(\frac{-e_l\prime(x)}{e_l(x)})\right]^{\prime}
\nonumber \\
=\frac{(n-1)\pi\nu^2}{8}\left[1-
\frac{41}{128}\frac{1}{\nu^2}+\frac{711}{16384}\frac{1}{\nu^4}\right]- \nonumber
\\
\frac{3(n-1)^2\pi\nu^2}{32}\left[1-\frac{592}{315}\frac{1}{\pi\nu}-
\frac{33}{128}\frac{1}{\nu^2}+ \right. \nonumber \\
\left.
\frac{29504}{45045}\frac{1}{\pi\nu^3}+\frac{639}{16384}\frac{1}{\nu^4}\right],
\end{eqnarray}
and the exterior nondispersive surface force density $F_{ext}^{nondisp}$    is
\begin{equation}
F_{ext}^{nondisp}=-\frac{1}{4 \pi^2
a^4}\sum_{l=1}^{\infty}\int_0^{\infty}dz\nu^3\! z\Lambda_l^{ext}
[\ln(\frac{-e_l\prime(x)}{e_l(x)})]^{\prime}.
\end{equation}
 When constructing the total nondispersive surface force density as
$F^{nondisp}=F_{int}^{nondisp}
+F_{ext}^{nondisp}$,  we see from (20) and (22) that the terms of order $(n-1)$
{\it compensate each other}.
 The surface force accordingly starts with terms containing $(n-1)^2$.

(The application of the general formalism developed in \cite{brevik94} to dilute
balls contains a calculational
error; Eq.(94) should have been multiplied by a factor 1/2. Therefore the final
conclusion of Sect.
4 in \cite{brevik94} is incorrect. We thank K. A. Milton for informing us about
this point.)

  Adding (21) and (23) we now get
\begin{equation}
F^{nondisp}= \frac{(n-1)^2}{64 \pi a^4}\sum_{l=1}^{\infty}
(\nu^2-\frac{65}{128}+\frac{927}{16384}\frac{1}{\nu^2})
\end{equation}
Here the divergence mentioned above is made explicit; regularization is needed
in the first two
terms. We shall use the Riemann zeta function regularization. This method has in
recent years become a standard tool in all areas of quantum field theory, and is
found to be in agreement with what one can derive by other means, such as the
dimensional regularization, cutoff regularization, etc. (cf, for instance, the
monograph of Elizalde {\it et al.} \cite {elizalde94}). Zeta function
regularization has been applied in
the context of spherical dielectrics in \cite{brevik94} and \cite{brevik90}.
An interesting  contribution to this technique is further provided by the recent
paper of Lambiase {\it et al.} \cite{lambiase98}. Here, no formal divergences
appear; an extra parameter $s$ is introduced whereby all expressions are
retained finite. At the end, an analytical continuation of $s$
is performed. The final results are in agreement with what one  finds by a
straightforward application of the Riemann zeta function technique.

To effectuate this kind of regularization we need in practice only the formula
\begin{equation}
\sum_{l=0}^{\infty}\nu^s=(2^{-s}-1)\zeta(-s),
\end{equation}
according to which
\begin{equation}
\sum_{l=1}^{\infty}\nu^2=-\frac{1}{4},~~~~
\sum_{l=1}^{\infty}\nu^0=-1.
\end{equation}
The third sum in (24) is finite and is known exactly:
$ \sum_{l=1}^{\infty}\nu^{-2}=(\pi^2/2)-4 . $
 From (24) we then
get, when writing the force as $ F^{nondisp}|_{1/\nu^4} $   to emphasize that it
is based upon a uniform asymptotic expansion
to order $ 1/\nu^4 $:
\begin{equation}
F^{nondisp}|_{1/\nu^4}=\frac{(n-1)^2}{4 \pi a^4}
[\frac{129}{65536}+\frac{927}{524288}\pi^2 ].
\end{equation}
It is of interest to compare this expression with the expression
$ F^{nondisp}|_{1/\nu^2} $ following from a uniform asymptotic expansion to
order $ 1/\nu^2 $ only:
\begin{equation}
F^{nondisp}|_{1/\nu^2}=\frac{(n-1)^2}{4 \pi a^4}
\frac{33}{2048}.
\end{equation}
The reason for this comparison is the following: the combination of terms
occurring in the surface force density (24) involves even powers of $ 1/\nu $
only. The structure of terms is such that the $ O(1/\nu^4) $ force is somewhat
larger than the correct force, whereas the $ O(1/\nu^2) $ force is somewhat
smaller. From (27) and (28) we are therefore able to give quite accurate bounds
for the "exact" nondispersive surface force density $ F^{nondisp} $ :
\begin{equation}
F^{nondisp}|_{1/\nu^2}<F^{nondisp}<F^{nondisp}|_{1/\nu^4}.
\end{equation}
(The term "exact" here refers to the uniform asymptotic expansion only; we are
of course concerned with a dilute-medium theory of order
$ (n-1)^2 $.) Numerically,
\begin{equation}
\frac{(n-1)^2}{4 \pi a^4} 0.0161<F^{nondisp}<
\frac{(n-1)^2}{4 \pi a^4}0.0194.
\end{equation}
The nondispersive part of the surface force is thus {\it repulsive}. The quantum
mechanical result is in this sense in agreement with the prediction of the
Abraham-Minkowski term in (1).

We note that some care has to be taken to include a sufficient number of terms
of the uniform asymptotic  expansion when constructing the expression for $
F^{nondisp} $. To illustrate this, let us see what becomes the result from
including only the {\it first} term in (24) (this corresponds to including only
the zeroth order of the uniform asymptotic expansion). Making use of the first
regularizing equation in (26) we obtain
\begin{equation}
F^{nondisp}|_{1/\nu^0}=-\frac{(n-1)^2}{4 \pi a^4}
\frac{1}{64}.
\end{equation}
Since 1/64 = 0.0156 we see that the magnitude of (31) is not very different from
the second order approximation in (30), but the {\it sign} is wrong. We conclude
that it is necessary to go to at least the second order in the uniform
asymptotic expansion to get the sign right. The expression (31) is actually in
precise agreement with the nondispersive force as calculated by Milton
\cite{milton80},
\cite{milton96}, and Milton and Ng \cite{milton97}. The reason why these authors
got a discrepancy with (30) is simply that they included an insufficient number
of terms . In \cite{brevik98} and \cite{milton99} the following accurate value
is given, with the assistance of numerical methods,
\begin{equation}
F^{nondisp}=\frac{(n-1)^2}{4 \pi a^4}0.01907
\end{equation}
The fourth order result given in (27) is about 1.7 per cent high.

\section{ Dispersive force}

This case is more difficult to handle than the nondispersive case, and we will
essentially be able to give only order-of-magnitude estimates. We restrict
henceforth the uniform asymptotic
expansion to the second order. Inclusion of higher order  terms would here be
pointless.
The ball is assumed to be dilute, as before.

  We now have to keep the nondimensional ultraviolet cutoff $x_0=\omega_0 a $ as
a finite quantity in thr formalism.  One may ask: what are the physically
reasonable magnitudes of  $ x_0 $ ?  We note that $ x_0 $ is here an "external"
parameter that has to be inserted into the formalism by hand. Its
value is given by the cutoff frequency $\omega_0 $ , which is determined by the
molecular structure of the
medium, and by the radius $ a $. We will take $ \omega_0 $  to lie in the
ultraviolet region. It seems reasonable,
by comparison with the Lorentz dispersion model (cf. \cite{schwinger78}, for
instance), to put $ \omega_0 = 3\times 10^{16} s^{-1} $.
Thus in dimensional units, assuming a small ball with radius $ a = 1 \mu m $, we
get $ x_0 = \omega_0 a/c = 100 $.
Balls with somewhat larger radii can also be actual. Accordingly,
 \[ 100 \leq x_0 \leq 1000 \]
seems to be a reasonable range for $ x_0 $.

     Another question is: what is the upper physical limit $ l=l_0 $ to be taken
in the
summations? This point becomes accentuated by the circumstance that some of the
sums are
formally divergent. Obviously, this behaviour is a spurious effect. Our
dispersion relation, as
explained in the beginning of Sect. 2, implies that photons having frequencies $
\omega $   higher than $ \omega_0 $
 do not "see" the medium at all. If a photon of limiting frequency $ \omega_0 $
just touches the surface
of the sphere, its  angular momentum is equal to $ \omega_0 a $  , i.e., equal
to $ x_0 $.  We thus expect, on { \it physical}
grounds, that $ l_0 $   is of the same order as $ x_0 $. This kind of argument
has repeatedly been used in
Casimir calculations  \cite{candelas82}, \cite{brevik90}, \cite{brevik94a}. It
means that one confines oneself to giving order-of-magnitude
estimates rather than exact numbers. The physical force, of course, is a
definite number in a given
case, but to evaluate it one needs detailed information about the dispersive
structure of the
medium. That lies outside the scope of the present paper.

     Adding (15) and (17), making use of the second order approximation, we get
\begin{eqnarray}
F &=& F_{int}+F_{ext} \nonumber \\
&&= \frac{(n-1)^2}{4 \pi^2 a^4}\sum_{l=1}^{\infty}
\nu^2 [ \int_0^{x_0/\nu}z^2t^6(z)dz \nonumber \\
&& + \frac{1}{4 \nu^2}\int_0^{x_0/\nu}
z^2t^{12}(z)(3-42z^2+18z^4+2z^6)dz+ O(\frac{1}{\nu^4})
\end{eqnarray}
Here the first and the second integral stem respectively from the zeroth, and
the second, order
terms in the uniform asymptotic expansion.  We write $F $ as a sum of two terms,
one zeroth order term $ A $ and one second order term $  B $, and consider $B$
first. This term is more easy to handle than the term $A$, since the sums over
$l$ in $B$ are convergent and can be calculated by means of the Euler-Maclaurin
formula \cite{abramowitz64}. It is convenient to write $B$ in the form
\begin{equation}
B= \frac{(n-1)^2}{4 \pi^2 a^4}\frac{1}{4}
\sum_{l=1}^{\infty}\int_0^{x_0}x^2
\frac{3\nu^9-42x^2\nu^7+18x^4\nu^5-2x^6\nu^3}
{(x^2+\nu^2)^6},
\end{equation}
 and to use the Euler-Maclaurin formula in the following version:
\begin{eqnarray}
\sum_{l=1}^{\infty}f(l)&=&\sum_{l=1}^{L} f(l) +\int_{L+1}^{\infty}dl f(l)
\nonumber\\
&&+\frac{1}{2}[f(\infty)+f(L+1)]+
\frac{1}{12}[f'(\infty)-f'(L+1)]+...
\end{eqnarray}
Here $L$ is an auxiliary integer, the use of which may improve the accuracy of
the series. Usually it suffices to let the magnitude of $L$ be moderate ($L\leq
10$; cf. \cite{brevik90} for instance, where we made an analogous calculation).
We write the expression (34) as $B=\sum_{l=1}^{\infty}f(l)$, and obtain for the
first term in (35) to a good accuracy
\begin{multline}
\sum_{l=1}^Lf(l)=\frac{(n-1)^2}{4 \pi^2 a^4}\frac{L}{4}
\int_0^{\infty}z^2t^{12}(z)(3-42z^2+18z^4-2z^6)dz = \\
= -\frac{(n-1)^2}{4 \pi a^4}\frac{65}{2048}L.
\end{multline}
Here we could set the upper limit of the integral equal to infinity, since the
main contribution comes from $z$ close to the lower limit. The second term in
(35) yields after some algebra, when introducing the abbreviation $ w= L+3/2 $,
\begin{eqnarray}
\int_{L+1}^{\infty}dlf(l)\nonumber\\
&=&\frac{(n-1)^2}{4 \pi^2 a^4}[-\frac{x_0}{8}+
\frac{1}{64}\frac{w^8x_0}{(w^2+x_0^2)^4}\nonumber\\
&&-\frac{11}{128}\frac{w^6x_0}{(w^2+x_0^2)^3}+
\frac{105}{512}\frac{w^4x_0}{(w^2+x_0^2)^2}\nonumber\\
&&-\frac{325}{1024}\frac{w^2 x_0}{w^2+x_0^2}
+\frac{315}{1024}w\arctan\frac{x_0}{w}-\frac{125}{1024}w\pi]\nonumber\\
&& \rightarrow \frac{(n-1)^2}{4 \pi^2 a^4}
(-\frac{x_0}{8}+\frac{65}{2048}w\pi), ~~~x_0 \rightarrow \infty.
\end{eqnarray}
The last term between square parentheses in (37) was evaluated putting $x_0=
\infty$. Similarly we put $x_0=\infty$ when evaluating the last nonvanishing
terms in (35):
\begin{equation}
\frac{1}{2}f(L+1)-\frac{1}{12}f'(L+1)=
-\frac{(n-1)^2}{4 \pi^2 a^4}\frac{65}{4096}\pi.
\end{equation}
Altogether, for large cutoffs, we get the $L$-independent result
\begin{equation}
B=\frac{(n-1)^2}{4 \pi^2 a^4}(-\frac{x_0}{8}+
\frac{65}{2048}\pi),~~~~x_0\rightarrow \infty.
\end{equation}
We extract the nondispersive part of this expression:
\begin{equation}
B( nondispersive~~  part )=\frac{(n-1)^2}{4 \pi a^4}
\frac{65}{2048}.
\end{equation}
This is compared with the second order contribution to the nondispersive force
obtained previously in Eq.(24):
\begin{equation}
F^{nondisp}(2nd~~ order)=\frac{(n-1)^2}{64 \pi a^4}
(-\frac{65}{128})\sum_{l=1}^{\infty}\nu^0=
\frac{(n-1)^2}{4 \pi a^4}\frac{65}{2048}.
\end{equation}
The expressions (40) and (41) are seen to be equal. This brings us to the
following important conclusion: If we perform a {\it dispersive} calculation of
the surface force, to second order, we find a term that is independent of the
cutoff. This term is precisely equal to the second order contribution  within a
calculation that is {\it nondispersive} from the outset. In the second case
zeta-function regularization is being employed, while in the first case it is
not.

Consider next the zeroth order term $A$ in (33). Performing the integration over
$z$ we get
\begin{eqnarray}
A\equiv && \frac{(n-1)^2}{4 \pi^2 a^4}\sum_{l=1}^{\infty}
\nu^2\int_0^{x_0/\nu}z^2t^6(z)dz \nonumber\\
&=&\frac{(n-1)^2}{4 \pi^2 a^4}\sum_{l=1}^{\infty}
[-\frac{1}{4}\frac{\nu^5 x_0}{(\nu^2+x_0^2)^2}+
\frac{1}{8}\frac{\nu^3 x_0}{\nu^2+x_0^2}+
\frac{\nu^2}{8}\arctan \frac{x_0}{\nu}].
\end{eqnarray}
We will only be able to give an order-of-magnitude analysis of the physically
meaningful part of this divergent expression. Note first, however, the following
point: take the nondispersive limit, $x_0=\infty $. Then only the last term in
(42) survives and we get, again employing zeta function regularization,
\begin{equation}
A( nondispersive~~ part)=\frac{(n-1)^2}{4 \pi a^4}
\frac{1}{16}\sum_{l=1}^{\infty}\nu^2=
-\frac{(n-1)^2}{4 \pi a^4}\frac{1}{64}.
\end{equation}
This is precisely the same as the expression (31), and serves as a corollary of
our dispersive calculation.

Now return to the dispersive case, and estimate the contribution to the sum in
(42) from low and moderate values of $l$, such that $\nu\leq x_0$. When $\nu\ll
x_0$, we see that the two first terms in (42) are negligible in comparison to
the third term. Moreover, when $\nu=x_0$ the two first terms compensate each
other. Consequently we may write
\begin{equation}
A|_{\nu\leq x_0}\simeq\frac{(n-1)^2}{4 \pi^2 a^4}
\sum_{\nu\leq x_0}\frac{\nu^2}{8}\arctan \frac{x_0}{\nu}.
\end{equation}
Here the variation in the factor $ \arctan(x_0/\nu) $ is moderate; it decreases
from  $\pi/2$ for small $\nu$ , to
$ \pi/4$  when $x_0 =\nu$ . For estimate purposes we may replace
$\arctan(x_0/\nu )$ by a constant $C$, where $C$
lies in the interval $\pi/4\leq C \leq \pi/2$.  The sum in (44) then becomes
simple to evaluate, since
\begin{equation}
\sum_{l=1}^{x_0-1/2}\nu^2=\frac{1}{3}x_0^3+\frac{1}{2}
x_0^2+\frac{1}{6}x_0-\frac{1}{4},
\end{equation}
and so we get
\begin{equation}
A|_{\nu\leq x_0}=\frac{(n-1)^2}{4 \pi^2 a^4}
\frac{C}{8}(\frac{1}{3}x_0^3+\frac{1}{2}x_0^2
+\frac{1}{6}x_0-\frac{1}{4}).
\end{equation}
It is here natural to argue such as we did above, in connection with (40) and
(41). We may thus associate
the $x_0 $- independent part of (46) with the nondispersive surface
force, calculated according to the zeroth order uniform asymptotic expansion.
Actually, if we equate the $ x_0$ -
independent part (arising from the last term in (46)) to the expression (31), we
obtain $C =  \pi/2$.
This is precisely the value of $C$ corresponding to the {\it small} values of
$\nu$   , according to our
discussion above. We may thus conclude that the nondispersive part of the force,
at least as far
as the leading term is concerned, is associated with the low angular momenta $l$
.

     When $x_0$ is included, (46) predicts under usual physical circumstances a
{\it strong, repulsive}
force. For instance, if we take  $C =\pi /2$  and  $ x_0 = 100$, we find that
(46) is about  $10^6$   as large
as the nondispersive second order (or fourth order) force given in (30). This
large difference
can be expected to be quite important, in practice. An eclatant example to think
about is the
sonoluminescence effect \cite{milton97}. We will briefly return to this point in
the Conclusion section.

The result (46) was derived under the assumption of low and moderate
nondimensional frequencies, $\nu\leq x_0$. Is there an appreciable contribution
to $A$ from higher frequencies also? We think that this is not so, the reason
being, as explained above, that photons with $\omega > \omega_0$ do not "see"
the medium. It is physically speaking most safe to truncate the $l$ summation at
$\nu \simeq x_0$ and adopt (46) as a reasonable order-of-magnitude estimate of
the leading dispersive term. The total surface force density
$F$ is thus estimated adding the expressions (46) and (39).

Finally we mention the following alternative method of handling the divergent
sum $A$, which may appear mathematically convenient. The method consists in
adding and subtracting the weakly divergent sum representing $A$ in the
frequency region $\nu > x_0$. Starting from (42) we see that this amounts to
writing
\begin{multline}
A=\frac{(n-1)^2}{4 \pi^2 a^4}\sum_{l=1}^{\infty}
[-\frac{1}{4}\frac{\nu^5 x_0}{(\nu^2+x_0^2)^2}+
\frac{1}{8}\frac{\nu^3 x_0}{\nu^2+x_0^2}+
\frac{\nu^2}{8}\arctan \frac{x_0}{\nu}-\frac{x_0^3}{3\nu}]
+\\ +\frac{(n-1)^2}{4 \pi^2 a^4}\frac{x_0^3}{3}
\sum_{l=1}^{\infty}\frac{1}{\nu},
\end{multline}
where we have chosen to let the sums over $l$ in the extra terms run from $1$ to
infinity. The advantage of writing $A$ in this way is that the first of the two
sums in (47) is finite, and that we have explicit control over the divergence in
the second sum. Let the first sum be denoted by $ A({\rm finite})$. Making a
numerical calculation of this expression at the lower and upper limits $x_0=100$
and $x_0=1000$, we get approximately
\begin{equation}
A( finite)\simeq \frac{(n-1)^2}{4 \pi^2 a^4}
\left\{\begin{array}{ll}
-1.7\times 10^6,~~~~x_0=100\\
-2.4\times 10^9,~~~~x_0=1000
\end{array}
\right.
\end{equation}
A graphical representation of $A({\rm finite})$ versus $x_0$ shows that the
variation is roughly linear in the actual frequency region. Exploiting this
fact, and truncating the second sum in (47) at $ l=x_0 $, we arrive at the
following estimate for $A$:
\begin{equation}
A\simeq \frac{(n-1)^2}{4 \pi^2 a^4}\left[\frac{x_0^3}{3}
\sum_{l=1}^{x_0}\frac{1}{\nu}-(266.5\frac{x_0}{100}
-264.8)10^6\right],
\end{equation}
valid for $100\leq x_0\leq 1000 $. It is seen, in fact, that the structure of
(49) is not very different from the structure of our previous estimate (46). The
advantage of (49) in comparison with (46), of course, is that there appears no
extra constant $C$ to be estimated separately.

\section{Conclusion and final remarks}

     The physical model considered in this paper is that of a compact dilute
spherical ball, surrounded
by a vacuum. The dispersion relation of the material is such that $n(i
\hat\omega)=n=const$   when the
imaginary frequency $\hat\omega$   is lower than the ultraviolet cutoff
frequency $\omega_0$   whereas
$n(i \hat\omega)=1$ for higher $\hat\omega$ . This is a very simple dispersion
relation: we know on general grounds that $n(i \hat \omega)$, being closely
related to a so-called generalized susceptibility, is real on the positive
imaginary frequency axis and decreases monotonically towards 1 when $\hat \omega
\rightarrow \infty$
(cf., for instance, \cite{landau80}). However, this is not a serious restriction
here as we may well imagine that  $n(i \hat \omega)$ instead is endowed with a
small negative slope for $\hat\omega < \omega_0$.

     The nondimensional cutoff is $x_0=\omega_0 a$,  $ a $ being the radius of
the ball. Under typical
physical conditions, if the radius lies in the interval  $1 \mu m \leq a \leq 10
\mu m $,  we expect that $
 100\leq x_0 \leq 1000 $.

     Our basic Green function formalism (Sect. 2) is Taylor expanded in $(n-1)$
up to the
second order, and is expanded in $1/\nu$  up to the fourth order  ($\nu  = l  +
1/2$). The Maple V program
here proves to be an effective tool.

     The surface force density $F$ is constructed by taking the difference
between the radial
Maxwell stress tensor components on the two sides of the boundary  $ r = a$ .
Alternatively, $F$ can
be found as the integral of the radial component of the volume force density
${\bf f}$  across the
boundary. The general expression for ${\bf f} $  is given in (1), and the
meaning of the various terms
is discussed. Only the first, and under usual conditions most important, term
(the Abraham -
Minkowski term) is further considered in this paper

The {\it nondispersive} force is calculated in Sect.3. Formally, this case is
treated by setting $x_0=\infty$, and regularizing the divergent sums using the
Riemann zeta function. Working to the order $1/\nu^4$ in the uniform asymptotic
expansion, the nondispersive force is given by (27) and (30). A slightly more
accurate value is given by (32). This force is repulsive.

We emphasize that $F^{nondisp}$ is a {\it physical}  force; the expression is
not merely a formal outcome of a more or less arbitrary regularization
procedure.  It is here of interest to note that, in addition to the Green
function approach above, there has recently been developed also two other
alternative procedures which lead to results  supporting our statement:

First, one may consider the problem from a microscopic point of view and
calculate the mutual van der Waals energy for the molecules in the compact
sphere. This has been done by Milton and Ng \cite{milton98}. The van der Waals
energy for two molecules is $V=-B/r^7 $, with $B$ a positive constant.
Integrating this over the sphere one finds a divergent expression which has to
be regularized, conveniently by means of dimensional regularization. The finite
remaining part of the energy becomes
\begin{equation}
E=\frac{(\varepsilon-1)^2}{\pi a}\frac{23}{1536}
\end{equation}
Associating the van der Waals energy $E$ with the surface force $F$ via the
relation
$ F=-(1/4 \pi a^2)\partial E/\partial a $, we find the same numerical result for
$F=F^{nondisp}$ as in Eq. (32).

Second, one can calculate the mutual energy between the molecules in the sphere
by employing quantum mechanical perturbation theory. This has been done by
Barton \cite{barton98}, for a dilute sphere. He used an exponential cutoff for
the wave numbers. Again, the formal expression for $E$ was found to diverge, but
after omission of the cutoff dependent terms, the result was found to be
precisely in agreement with the expression (50).

These two recent alternative developments are welcome, since they signify that a
consensus between various approaches to the difficult Casimir surface force
problem is finally in sight.

We emphasize that the repulsive force $F^{nondisp}$, although being a physical
force and thus measurable in principle, is {\it not} the total surface force,
not even for a nondispersive medium. The total force is necessarily {\it
attractive}, since it results from the attractive van der Waals force between
molecules. (This point was apparently regnogized very early, by Davies in his
classic paper \cite{davies72}.) At first sight it may perhaps be surprising that
we operate with a measurable force component which is only a part of the total
force. This kind of situation is however known also from other areas in
electrodynamics, and it by be instructive in the present context  to discuss
briefly the following two examples:

1)  Consider the situation where a high intensity laser beam falls vertically
from above on a liquid (water) surface. The surface bulges {\it out},
corresponding to an outward-directed force. The elevation is slight, of the
order of $1~ \mu m$, but is clearly detectable. This is the famous Ashkin-
Dziedzic experiment \cite{ashkin73}. The elevation can be described completely,
using only the Abraham-Minkowski force term, {\it i.e.,} the first term in (1).

The Abraham-Minkowski force is however not the only force acting on the water
surface. There is an additional force, arising from electrostriction, and this
force is stronger than the first-mentioned one such that the total force on the
surface is actually {\it compressive}. Electrostriction does not contribute to
the elevating force, as mentioned earlier, since electrostriction merely
compresses the column of liquid. The theory of the Ashkin-Dziedcic experiment
has been given in \cite{lai76} and \cite{brevik79}.

2)  A second, related, situation is the one discussed in many textbooks ( cf.
\cite{landau84}, for instance), namely two parallel condenser plates partly
immersed in a dielectric liquid. When a strong  static electric field is applied
between the plates, the liquid is observed to rise slightly in the interior
region. The situation is more involved than one might be inclined to think.
Again, the observable elevation can be descibed entirely with use of the
Abraham-Minkowski force only. There is in addition the compressive
electrostriction force acting, so that the total surface force density is
compressive. The electrostriction force has to be the stronger member of the two
forces, since otherwise the column of liquid would not be kept together as a
whole. The mathematical reason why the electrostriction force is the stronger
one, is that the relative permittivity is larger than 1. This is a
thermodynamical result. Seen from this perspective, the deepest reason why the
column of liquid is kept together, is neither hydrodynamics nor electrodynamics,
but rather {\it thermodynamics}. The theory of this situation is discussed,
among other places,  in \cite{brevik79} and \cite{brevik82a}.

The lesson to be learned from the above two examples is fairly obvious: it is
physically quite legitimate to consider a part of the complete surface force,
and discuss its experimental consequences, without drawing the complete force
into consideration at all. This is precisely what we do in principle  when we
discuss the Casimir force $F^{nondisp}$. We are not aware of any concrete
operational argument so far to really measure $F^{nondisp}$ experimentally,
however.

Consider next the dispersive force. It was calculated in Sect. 4, to second
order in the asymptotic expansion.  The force is written as a sum of two terms,
$F=A+B$, where the second order term $B$ is given in (39). This term, although
strictly speaking derived for high cutoffs only, $x_0\rightarrow \infty$, is
expected to be fairly accurate for all values of $x_0$ that are actual in
practice. The term $B$ has the same basic structure as the surface force
calculated earlier on a spherical {\it shell} \cite{brevik90}: there is an
attractive part linear in $x_0$, which is due to dispersion, and in addition a
repulsive part which is dispersion independent. Under usual physical conditions
the first of these two parts is the stronger one.

As for the leading term $A$, of zeroth order, we have only been able to give an
order-of-magnitude analysis, based upon physical arguments allowing us to
truncate the divergent sums over $l$ at an upper limit approximately equal to
$x_0$. We gave two estimates for $A$, in equations (46) and (49). The structures
of these two expressions are seen to be roughly similar. Under usual
circumstances also the force $A$ is a repulsive one.

To get an idea about the magnitudes involved, let us limit ourselves to the
dominant first term in (46), putting for definiteness $C=\pi/2$. Then
\begin{equation}
F^{disp} \simeq \frac{(n-1)^2}{4 \pi a^4}\frac{x_0^3}
{48}=\frac{(n-1)^2}{4 \pi a}\frac{\omega_0^3}{48}.
\end{equation}
This force may easily be about $10^6$ as strong as the nondispersive force (27).
The force (50) corresponds to the Casimir energy (integrating while keeping
$\omega_0$ constant)
\begin{equation}
E^{disp} \simeq -\frac{(n-1)^2\omega_0^3}{96}a^2.
\end{equation}
This energy is always negative, and has its maximum when the radius is at
minimum. Moreover it is independent of the substitution $n\rightarrow 1/n$, when
$n$ is close to unity. One would therefore think that the formula is equally
well applicable to the case when there is a vacuum bubble immersed in a dilute
dielectric medium.

One may wonder whether it is possible to apply the present kind of theory to the
phenomenon of sonoluminescence (an instructive review of this phenomenon has
recently been given by Cheeke \cite{cheeke97}. It is well known that the
energies involved in the nondispersive Casimir effect are far too small to
account for the sonoluminescence effect. Can the very large magnitude of (52)
"save" the situation? We see that by taking $(n-1)^2 \simeq 0.1, ~~x_0 \simeq
1000$ (corresponding to $a \simeq 10 \mu m$), we obtain from
(52) $E^{disp}\simeq -2\times 10^4 eV$. The magnitude of this energy is probably
acceptable, but the sign is apparently wrong \cite{milton96}, \cite{milton97}.
At least the sign is in conflict with energy emission {\it during the collapse}.
Consider, however, the following argument: an experimental fact is that the
photon flash does not occur during the whole collapse period but instead
precisely at the end of that period, when $a=a_{min}\simeq 1 \mu m$. Imagine now
that the energy of the burst of photons were really taken from the Casimir
energy at this particular instant. Consequently there would occur an energy
imbalance in the system. The balance of Casimir energy would be restored if the
bubble were increasing its radius out to $a\simeq 10 \mu m$.

Such a movement actually happens. Evidently, we do not "explain" the
sonoluminescence effect by this kind of argument. Our aim is more modest, viz.
to suggest that sonoluminescence is perhaps not totally unrelated to the Casimir
effect after all.

The last point that we shall dwell on in this paper, is the influence from {\it
absorption}  in the medium. This is an important point since it shows the
physical limitations of our adopted material model, and the directions for
further research. At first sight, it would seem as if absorption does not create
any problems at all in the theory: we are after all adopting a model in which
the quantity $n(i \hat \omega)$ is real and shows a simple though physically
acceptable behaviour along the positive imaginary frequency axis. Absorption
should accordingly be taken into account implicitly in the theory,  through the
Kramers-Kronig relations, which lead to a complex behaviour of $n(\omega)$ along
the real frequency axis. The problem waiting for a complete solution lies at
another place, namely in our adoption of formulas (7) and (8) for the {\it
effective products}. These are the same formulas as for a nondispersive medium.
A more complete theory would have to take into account the coupling of the
molecules to a {\it heat reservoir}, with a corresponding quantization of the
source currents. This would in turn imply the presence of {\it Langevin terms}
in the effective products. Interesting developments of the theory of absorptive
media, implying a quantization of the complete system (medium plus reservoir)
have been given in \cite{huttner92}, \cite{knoll92}, and most recently, in
\cite{dung98} and \cite{scheel98}. We only point to this problem here, without
going into any detail. For the convenience of the reader, we summarize in
Appendix B the basic assumptions leading to our formulas for the effective
products.

\section{Acknowledgments}
Valery Marachevsky would like to thank  Iver Brevik for hospitality and
encouragement during his scholarship research project in Norway in 1997.  He thanks Dmitri Vassilevich for 
active help in the preparation of the project and for valuable discussions. He also thanks The Research
Council of Norway, The International Scholarship Scheme, for financial support of this
project. Iver Brevik acknowledges valuable discussions and correspondence with  Gabriel
Barton, Michael Bordag,  Kimball Milton, Vladimir Nesterenko, and Stefan Scheel.

\appendix
\section{Nondispersive force expanded to the fifth order in $(n-1)$}

     Our calculation of the nondispersive force in Sect.3  was based on a
dilute-ball
expansion up to order $(n-1)^2$, and on an asymptotic expansion up to order $
1/\nu^4$. A characteristic
property was found to be that terms of order $(n-1)$ in $F_{int}^{nondisp}$  and
$F_{ext}^{nondisp}$  compensated each
other, so that the leading term in
$F^{nondisp}=F_{int}^{nondisp}+F_{ext}^{nondisp}$  would go as $(n-1)^2$; cf.
(20) and
(22). One may ask: what happens if one makes a higher order expansion in $(n-
1)$, will then terms
containing odd powers in $(n-1)$ continue to compensate each other? We shall
investigate this
point, giving the expansions for $F_{int}^{nondisp}$   and $F_{ext}^{nondisp}$
up to order $(n-1)^5$. However, we will now
have to truncate the uniform asymptotic expansion at order $1/\nu^2$. (A
combined set of higher order expansions
in $(n-1)$, as well as $1/\nu$ , turned out to be beyond the capacity of our
Maple program.)

     We abstain here from giving the explicit expressions for
$F_{int}^{nondisp}$  and $F_{ext}^{nondisp}$, but write
down their sum:
\begin{eqnarray}
F^{nondisp}&=&\frac{(n-1)^2}{4 \pi a^4}[\frac{33}{2084}-(n-1)(\frac{99}{4096}
+\frac{11}{210}\frac{1}{\pi}] \nonumber\\
&&+(n-1)^2(\frac{1213}{65536}+\frac{11}{105}\frac{1}{\pi})-(n-
1)^3(\frac{785}{131072}
+\frac{29}{216}\frac{1}{\pi})].
\end{eqnarray}
The first term in this expression agrees with our previous second order result
(28). We see that odd powers in $(n-1)$ generally do {\it not} compensate each
other in the surface force. All powers in $(n-1)$ are present, with the
exception of the first, linear, term. The expansion given above  may be useful
also in cases when $(n-1)$ is not so small, perhaps up to $ (n-1)\simeq 0.3 -
0.4$.

\section{The effective products derived from statistical mechanics}

Assuming for definiteness Schwinger's source theory, we relate the electric
field components $E_i(x)$
to the polarization components $P_k(x')$ via a tensor $\Gamma_{ik}(x,x')$:
\begin{equation}
E_i(x)=\int d^4x'\,\Gamma_{ik}(x,x')\,P_k(x')
\end{equation}
Stationarity of the system means that $\Gamma_{ik}$ depends on time only through
the difference
${\tau}=t-t'$. Causality  implies that the integration over $t'$ is limited to
$t' \leq t$. From a statistical mechanical viewpoint, $\Gamma_{ik}(x,x')$ is a
generalized susceptibility \cite{landau80}.

Introduce the Fourier transform $\Gamma_{ik}({\bf r}, {\bf r'},\omega)$ via
\begin{equation}
\Gamma_{ik}(x,x')=\int_{-\infty}^{\infty}\frac{d\omega}{2 \pi}\,e^{-i\omega
\tau}
\, \Gamma_{ik}({\bf r},{\bf r'},\omega)
\end{equation}
The basic equation (54), in combination with the causality principle, are thus
the necessary conditions for causing the function $\Gamma_{ik}({\bf r}, {\bf
r'}, \omega)$ to satisfy the following properties:

{(\it i)} It is a regular, one-valued function in the upper half frequency
plane. At infinity $(|\omega|
\rightarrow \infty)$, it goes to zero.

{(\it ii)}  There is no singularity on the real axis  (except at the origin, in
the case of a metal).

{(\it iii)}  The function does not take real values at any finite point in the
upper half plane except on the imaginary axis.

Now invoke {\it Kubo's formula} from statistical mechanics: it states that the
spectral generalized susceptibility is related to the commutator between field
components as
\begin{equation}
\Gamma_{ik}({\bf r},{\bf r'},\omega)=i\int_0^{\infty}d\tau\,
e^{i\omega\tau}\,\langle
[E_i(x),E_k(x')]\rangle
\end{equation}
(cf. for instance, \cite{landau80}). This means that the generalized
susceptibility is identifiable with the  retarded Green function:
$\Gamma_{ik}(x,x')=G_{ik}^R(x,x')$. For $t<t'$ both $\Gamma_{ik}(x,x')$ and
$G_{ik}^R(x,x')$ vanish, the first because of {\it causality}, the second
because of the {\it definition} of the retarded Green function.

Consider now the two-point function $ \langle E_i(x)E_k(x')\rangle $. Its
Fourier transform
$\langle E_i({\bf r},\omega)E_k({\bf r'},\omega')\rangle $ can be expressed in
terms of the quantity
$ \langle E_i({\bf r}) E_k({\bf r'}) \rangle_{\omega} $  (the spectral
correlation tensor) as
\begin{equation}
\langle E_i({\bf r},\omega)E_k({\bf r'},\omega') \rangle=
2\pi\langle E_i({\bf r})E_k({\bf r'})\rangle _{\omega}\,\delta(\omega+\omega')
\end{equation}
The crucial step now is to apply the {\it fluctuation-dissipation theorem}
\cite{landau80}, according to which
\begin{equation}
\langle E_i({\bf r}) E_k({\bf r'})\rangle_{\omega}=Im\,G_{ik}^R({\bf r},{\bf
r'},\omega)\,
\coth (\frac{1}{2}\beta\omega),
\end{equation}
with $\beta=1/(k_BT)$. At zero temperature,
$\coth(\frac{1}{2}\beta\omega)\rightarrow sgn\,(\omega)$. Choosing the causal
integration contour in the complex frequency plane we obtain, after a complex
frequency rotation, the effective products as given in Eqs.(7) and (8).

\end{document}